\documentclass[twocolumn,nofootinbib,amsmath,prd,aps,superscriptaddress,
tightenlines,preprintnumbers,nobalancelastpage]{revtex4}

\usepackage{slashed}
\usepackage{color, verbatim}

\usepackage{latexsym}
\usepackage{amsmath}
\usepackage{graphicx}

\usepackage{amssymb}
\usepackage{graphicx}
\usepackage{bm}
\usepackage{adjustbox}

\usepackage{graphicx}
\usepackage{latexsym}
\usepackage{amsmath}
\usepackage{mathrsfs}
\usepackage{enumerate}
\usepackage{adjustbox}
\usepackage{amssymb}
\usepackage{slashed}

\def\bea{\begin{eqnarray}}
\def\eea{\end{eqnarray}}

\usepackage[normalem]{ulem}

\RequirePackage[colorlinks=true,urlcolor=blue,anchorcolor=blue,citecolor=blue,filecolor=blue,
               linkcolor=blue,menucolor=blue,linktocpage=true,pdfproducer=medialab,pagebackref=true]{hyperref}
\hypersetup{
  colorlinks   = true, 
  urlcolor     = mygreen, 
  linkcolor    = blue, 
  citecolor   = blue 
}
\usepackage{orcidlink}

 \usepackage[capitalise]{cleveref}
\usepackage{csquotes}
\definecolor{mygreen}{rgb}{0.0, 0.5, 0.0}

\usepackage[capitalise]{cleveref}

\usepackage[font=small,labelfont=bf,justification=centerlast]{caption}
\usepackage{mathtools}

\newcommand{\nn}{\nonumber}

\newcommand{\Mbf}{{\mathbf M}}
\renewcommand{\L}{{\mathcal L}}
\newcommand{\N}{{\mathcal N}}

\newcommand{\chiq}{\chi_{\text{QCD}}}
\newcommand{\agg}{{a}_{G\widetilde G}}
\newcommand{\pM}{\mbox{\large $p$}_{{\,\Mbf^2}}}
\newcommand{\pMone}{\mbox{\large $p$}_{{\,\Mbf_1^2}}}

\newcommand{\aPQ}{a_{\text{PQ}}}
\newcommand{\fPQ}{f_{\text{PQ}}}
\newcommand{\UPQ}{U(1)_{\text{PQ}}}
\renewcommand{\O}{{\mathcal O}}

\begin{document}

\newcount\hour \newcount\minute
\hour=\time \divide \hour by 60
\minute=\time
\count99=\hour \multiply \count99 by -60 \advance \minute by \count99
\newcommand{\mydate}{\ \today \ - \number\hour :00}

\preprint{}

\title{\Large The QCD axion sum rule} 

%
\author{Bel\'en Gavela\orcidlink{0000-0002-2321-9190}}
\email{belen.gavela@uam.es}
\affiliation{\it Departamento de F\'isica Te\'orica, Universidad Aut\'onoma de Madrid, \\ 
and IFT-UAM/CSIC, Madrid, Spain
}
\author{Pablo Qu\'ilez\orcidlink{0000-0002-4327-2706}}
\email{pquilez@ucsd.edu}
\affiliation{\it Department of Physics, University of California, San Diego, USA
}
\author{Maria Ramos\orcidlink{0000-0001-7743-7364}}
\email{maria.pestanadaluz@uam.es }
\affiliation{\it Departamento de F\'isica Te\'orica, Universidad Aut\'onoma de Madrid, \\ 
and IFT-UAM/CSIC, Madrid, Spain
}


\begin{abstract}
We demonstrate that the true QCD axion that solves the strong CP problem can be found in all generality outside the customary standard QCD band, with QCD being the sole source of Peccei-Quinn breaking. The essential reason is that the basis of axion-gluon interactions does not need to coincide with the mass basis.   
 Specifically, we consider the case in which
 the QCD axion field is not the only singlet scalar  in Nature but it mixes with other singlet scalars (besides the $\eta'$).  
 We determine the exact mathematical condition for an arbitrary  $N$-scalar potential to be  Peccei-Quinn invariant. 
Such potentials provide extra sources of mass for the customary axion without enlarging the Standard Model  gauge symmetry.
 The contribution to the axion mass stemming from the QCD topological susceptibility is shown to be shared then among the $N$ axion 
 eigenstates 
 through a precise sum rule. 
    Their location can only be displaced to the right of the standard QCD band.
  We demonstrate that the axion closest to this band 
  can be displaced from it by  a factor of $\sqrt{N}$ at most, 
  and this corresponds to the case in which all axion signals are maximally deviated.    
 Conversely, if one axion is found on the standard QCD band, the other eigenstates will be out of experimental reach.  
 Our results imply that any ALP experiment which finds a signal 
 to the right of  the standard QCD axion band can be solving the strong CP problem within QCD, with the associated $N-1$ excitations to be found in an area of parameter space that we determine.  
We illustrate the results and phenomenology in some particular cases.

\end{abstract}

\maketitle
\preprint{IFT-UAM/CSIC-23-58}


\newpage

\section{Introduction}
\vspace{-0.1cm}
Axions and axion-like particles (ALPs) are the subject of intense exploration at present, with both novel experimental proposals and theoretical work constantly being put forward.
Being pseudo-Goldstone bosons (pGBs), their search constitutes an epic and generic quest to uncover symmetries hidden in Nature and awaiting discovery.

The paradigmatic example  of this type of particle is the true axion that solves the strong CP problem of the Standard Model of Particle Physics (SM), that is, the fact that the value of one parameter of the strong force (QCD) - the $\bar \theta$ parameter- has to be adjusted by over ten orders of magnitude to comply with experimental bounds. In the successful wake of explaining small parameters through symmetries, the axion would be the pseudo-Goldstone boson of a dynamical solution via a global axial  $U(1)_A$ symmetry. This ``Peccei-Quinn'' (PQ)  symmetry must be classically exact although hidden (aka spontaneously broken), and explicitly broken  only by QCD quantum effects which give the axion $a$ a  tiny mass $m_a$\cite{Peccei:1977hh,Peccei:1977ur,Weinberg:1977ma,Wilczek:1977pj}. Indeed, the  anomalous  $a \,G_{\mu\nu}\tilde G^{\mu\nu}$ coupling to the gluon field strength $G_{\mu\nu}$
 is directly responsible for its mass $m_a$, which is then necessarily linked to the axion scale $f_a$ 
 by the relation\footnote{Which takes into account the mixing with the $\eta'$ and subsequently the $\pi^0$.}
 \begin{equation}
m_a\, f_a 
= \sqrt{\chi_{\rm QCD}} \simeq m_\pi \, f_\pi\,\frac{\sqrt{m_u\,m_d}}{m_u+m_d}  \,,
\label{invisiblesaxion}
\end{equation}
where $\chi_{\rm QCD}, m_\pi, f_\pi, m_u$ and $m_d$ denote respectively the QCD topological susceptibility, the pion mass, its decay constant, and the up  and down quark masses.  

Equation~(\ref{invisiblesaxion}) 
   defines the ``canonical QCD axion'', also often called the ``invisible axion'' or ``standard axion''. It corresponds to a straight line in the $\{m_a, 1/f_a\}$ parameter space: the QCD axion band. Its importance 
   stems from its presumed universality, i.e. it is presumed to hold for whatever  
    ultraviolet (UV) axion model, as far as QCD 
is the only source of Peccei-Quinn breaking.
 What lies beyond that band is considered to be ALP territory. ALPs  are pGBs similar to axions but {\it a priori} they are not associated to a solution of  the strong CP problem. Indeed the interest in axions extends far beyond the QCD axion; ALPs arise in a variety of theories (e.g. ~\cite{Dienes:1999gw,Gelmini:1980re,Davidson:1981zd,Wilczek:1982rv,Svrcek:2006yi,Arvanitaki:2009fg,Cicoli:2013ana,Alexander:2023wgk}) and are also attractive dark-matter candidates \cite{Preskill:1982cy,Abbott:1982af,Dine:1982ah}. In practice, the mass and scale of generic ALPs  do not need to abide by  Eq.~(\ref{invisiblesaxion}), and they are treated as independent free parameters.

  If an ALP signal  is  ever detected, an immediate and compelling question will be whether it can nevertheless be interpreted as a true axion that solves the strong CP problem.  Indeed, much theoretical effort is being dedicated to modify the relation in   Eq.~(\ref{invisiblesaxion}) so as to obtain for instance a heavier-than-QCD or lighter-than-QCD true axion.
   In other words, to obtain solutions in which the QCD band is displaced significantly towards the right~\cite{rubakov:1997vp,Berezhiani:2000gh,Gianfagna:2004je,Hsu:2004mf,Hook:2014cda,Fukuda:2015ana,Chiang:2016eav,Dimopoulos:2016lvn,Gherghetta:2016fhp,Kobakhidze:2016rwh,Agrawal:2017ksf,Agrawal:2017evu,Gaillard:2018xgk,Hook:2019qoh,Gherghetta:2020ofz,Csaki:2019vte,Kivel:2022emq} or the left ~\cite{Hook:2018jle,DiLuzio:2021pxd,DiLuzio:2021gos} of its standard position. This achievement requires in general   the enlargement the gauge group of the forces in Nature beyond the SM ones, providing new PQ violating contributions.
   
 In this paper we explore in depth a much simpler possibility.  
The physical axion mass eigenstate does not need to coincide
   with the scalar field that couples to $G_{\mu\nu}\tilde G^{\mu\nu}$ ({apart from the mixing} with the $\eta'$ and $\pi^0$ via QCD).  In other words, the axion-gluon interaction basis and the mass basis do not need to be simultaneously diagonal. Here, generic mixing of the axion 
   with extra scalars is allowed --alike to the mismatch of electroweak interaction versus mass bases for fermions. The axion field (or combination of fields) is thus allowed to mix with other $N-1$ arbitrary SM singlet scalars.
 As long as the mixing potential is classically PQ invariant, the solution to the strong CP problem holds. 
 The condition that has to be fulfilled by a general mixing potential  in order to be 
  PQ invariant will be identified.
It will be shown in all generality that the axion properties are shared among the $N$ eigenstates, with each of them obeying an equation which departs from  Eq.~(\ref{invisiblesaxion}) because the extra scalar potential provides extra contributions to their masses.  That is, one single axion field (or combination of fields) will reveal itself  in Nature as multiple axion signals. Specifically, 
  $N$ non-standard but true QCD axions will result, each of them with a different mass
$m_{i}$ 
 and different effective axion scale  $f_i$ 
 as determined from its coupling to gluons. 
 We will denote by $g_i$ the factor by which each of them 
  deviates from the standard relation,  
 \begin{align}
m_{{i}}^2f_{i}^2&\equiv{f_\pi^2 m_\pi^2} \frac{m_u m_d}{\left(m_u+m_d\right)^2} 
\times  g_i\,.
\label{Eq:gfactors}
\end{align}
    
 The system is tightly constrained, though, since the PQ symmetry dictates how
the mass induced by the topological susceptibility of QCD $\chi_{\rm QCD}$ is shared among the $N$ axion eigenstates. 
 Denoting by $\beta_i\equiv 1/g_i$ the  ``{\it axionness}''
  of each eigenstate 
$a_i$, 
 that is, the fraction of its mass which is due to the QCD topological susceptibility, it will be demonstrated that 
   \begin{equation}
  \sum_{i=1}^N\beta_i =1\,.
  \label{sumrule}
  \end{equation}
  This sum rule, together with other mathematically exact conditions and relations will be proven below in all generality.

  The constraint in Eq.~(\ref{sumrule}) is very rich in consequences, as it links the properties of a given axion eigenstate  to the fate of the other $N-1$ axions.
  Among other aspects of key importance for the experimental search, we will address and solve the question of what is the maximal possible displacement from the standard QCD band for  the axion eigenstate closest to that band.  The area in $\{m_a,1/f_a\}$ parameter space where these  multiple QCD axion  signals can be detected will be determined. 
  While for low values of $N$ the displacement for a given axion may be drowned on the error bands of the  projections, the prediction of up to $N$ different axion signals  is a direct smoking gun, for instance for experiments which measure the axion coupling to  the neutron electric dipole moment (nEDM)  operator  with very high precision in frequency, such as CASPER-electric \cite{Budker:2013hfa,JacksonKimball:2017elr},
 as the axion-to-nEDM coupling follows directly from the axion 
  gluonic coupling.
 
 The strength of other axion couplings to Standard Model (SM) fields is model-dependent. An important one is the axion-photon coupling, 
  explored in a plethora of experiments.  The modification of the latter  within the $N$ axion framework under discussion will be addressed as well and shown to also follow a displacement pattern, albeit subject to model-dependences.

 An underlying condition for the results in this paper to have strong experimental impact is that the contribution to the masses induced by the generic mixing potential should not be vastly different from the QCD-induced mass. Whenever this is not the case
 for some  axion eigenstates,   these will 
 be of no impact  (i.e. either too heavy or very light but in both cases decoupled from $G_{\mu\nu}\tilde G^{\mu\nu}$).
  That is, they will decouple from the sum rule in Eq.~(\ref{sumrule}) and thus will not significantly contribute to the solution to the strong CP problem.  In this perspective, the usual single canonical QCD axion is just one particular case of the vast parameter space of solutions, 
 in which all $N-1$ extra eigenstates have decoupled.
  
  A mixing of the canonical QCD axion with other singlet scalars in Nature has previously appeared in past publications~\cite{Kim:2004rp,Choi:2014rja,Kaplan:2015fuy,Giudice:2016yja,DiLuzio:2017ogq,Chen:2021hfq,Agrawal:2022lsp,Fraser:2019ojt,Darme:2020gyx}  but, either by choice or by construction, the limit in which all but one axion decouples was 
  taken and the features discovered here were not discussed.

The problem will be formulated within the model-independent framework of EFTs and for a generic scalar potential, and the main mathematical tool  to be used 
is the eigenvector-eigenvalue theorem~\cite{JacobiDeBQ,Denton:2019pka}.  Some UV complete toy examples will be also shown
 for illustration, though, in App.~\ref{App:UV}.

{
\section{Toy example: two axions}\label{sec:N2}

Before proceeding to formulate the problem for $N$ singlet scalars and the most general PQ-invariant potential, we illustrate some  results in this section within a $N=2$ EFT toy model. The reader interested in the general case and solutions can go directly to Sec.~\ref{multiple}. In what follows,  the notation $m_i$, $f_i$ will be reserved to denote the mass and the scale of a given mass eigenstate $a_i$, and $F$ will denote a certain combination of all $f_i$, see below.


Let us consider a combination of two fields with the following effective interactions:\footnote{The pattern of results to be obtained next will hold as well for a general mixing potential, i.e. $V(\hat a_1, \hat a_2)$ instead of the second term in this equation, see Sect.~\ref{multiple}.}

\begin{equation}
\label{eq:L2}
\mathcal{L}_{N=2} = \frac{\alpha_s}{8 \pi}\left(\frac{{\hat a}_1}{\hat f} + \frac{{\hat a}_2}{ \hat f} +\bar \theta\right) G \widetilde{G}-\frac{1}{2}\hat{m}_2^2 \, {\hat a}_2^2\,.
\end{equation}
Note that in this Lagrangian there is only one combination of fields that couples to $G \widetilde{G}$, ${{a}_{G\widetilde G}}/{F}\equiv  ({{\hat a}_1}+ {{\hat a}_2})/{\hat f}$, which is not necessarily a mass eigenstate, as it mixes with the orthogonal scalar combination, $a_{\perp}$, via a PQ invariant potential, i.e. 
\begin{equation}
\label{eq:L2-otro}
\mathcal{L}_{N=2} = \frac{\alpha_s}{8 \pi}\left(\frac{{a}_{G\widetilde G}}{F} +\bar \theta\right) G \widetilde{G}- 
\frac{1}{4}\hat{m}_2^2\, \big(a_{G\tilde G}-a_{\perp}\big)^2\,.
\end{equation}

It is easy to see 
that, in this particular example, 
it is the shift of $\hat a_1$ that implements the PQ symmetry and ensures CP conservation in the QCD vacuum. Expanding the axions around their minima we obtain the following potential,
\begin{equation}
\label{eq:V2}
V_{N=2} =  \frac{1}{2}\chi_{\rm QCD} {\left( \frac{{\hat a}_1}{\hat f} + \frac{{\hat a}_2}{\hat f}\right)^2} + \frac{1}{2}\hat{m}_2^2 \, {\hat a}_2^2
\end{equation}
which corresponds to the mass matrix,
\begin{equation}
\label{eq:MN2}
\mathbf{M^2} = \frac{\chi_{\rm QCD}}{\hat f^2}\begin{pmatrix}
1 & 1 \\
1 & 1 +r 
\end{pmatrix}\,,
\end{equation}
where $r \equiv \hat{m}_2^2 \hat{f}^2/\chi_{\rm QCD}$. Given the simplicity of the system it can be exactly diagonalized.
 The axion masses read 
\begin{equation}
m_{1,2}^2 = \frac{\chi_{\rm QCD}}{2 \hat f^2}\left(2  + r \mp \sqrt{4 + r^2}\right)
\end{equation}
and the corresponding physical eigenstates are given by
\begin{align}
&a_1 = \frac{{2 {\hat a}_1+{\hat a}_2\left(r-\sqrt{4+r^2}\right)}}{\sqrt{2} \sqrt{4+r^2-r \sqrt{4+r^2}}}\,,  \\
 &a_2 = \frac{{2 {\hat a}_2+{\hat a}_1\left(-r+\sqrt{4+r^2}\right)}}{\sqrt{2} \sqrt{4+r^2-r \sqrt{4+r^2}}}\,.
\end{align}
Generically, two axion mass eigenstates $a_1$ and $a_2$ result with different masses $m_1$ and $m_2$ and {\it both coupled to  $G \widetilde G$} with scales $f_1$ and $f_2$, 
\begin{align}
\mathcal{L}\supset\frac{\alpha_s}{8 \pi} \left(\frac{a_1}{f_{1}}+\frac{a_2}{f_{2}}\right) G\widetilde G \,.
\end{align}
For each eigenstate, we can also compute exactly the factor $g_i$ in \cref{Eq:gfactors},
\begin{align}
g_{1(2)} = \frac{2\sqrt{4 + r^2}}{\sqrt{4 + r^2} \pm \left(r-2\right)}\,,
\label{Eq:N=2g1g2}
\end{align} which describes how much the mass-decay constant relation deviates with respect to the single axion case, i.e. it measures the distance to the standard QCD band. For a single, canonical,  QCD axion $g=1$.

\begin{figure}[t]
\includegraphics[scale=0.4]{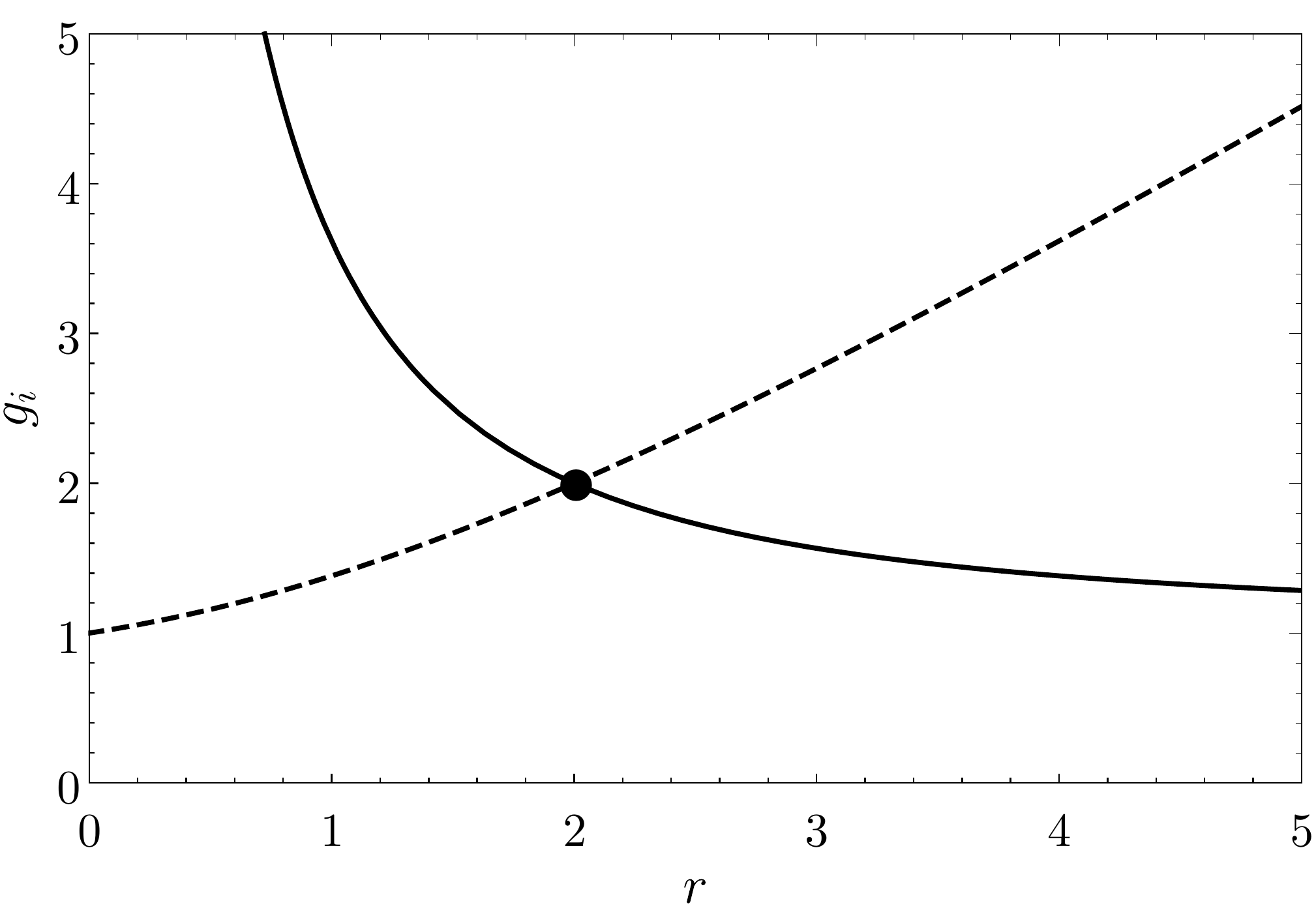}
\caption{The $g_i$-factors in \cref{Eq:N=2g1g2} as a function of $r$. For both limits, $r\rightarrow 0$ and $r\rightarrow \infty$, one axion decouples from the sum rule and the other is standard while for $r=2$ we obtain the maximal deviation and two~\textit{maxions} are generated.
{\hfill}
}
\label{fig:gN2}
\end{figure}
%

Now we are ready to study two widely known limits in which one of the axion eigenstates ends up behaving like a single QCD axion, i.e. either $g_1=1$ or $g_2=1$.
In the limit $r\to \infty$, one of the eigenstates, $a_2\simeq \hat a_2$,  becomes infinitely massive, and the other one, $a_1\simeq \hat a_1$, reduces to the standard single QCD axion.
In the opposite limit $r\to 0$, the determinant of $\mathbf{M}^2$ vanishes: one massive and standard QCD axion results $a_2\simeq (\hat a_1+\hat a_2)/\sqrt{2}$
,  while there is also a massless scalar but decoupled from the QCD sector $a_1\simeq (\hat a_1-\hat a_2)/\sqrt{2}$. 
For intermediate $r$, however, both axion eigenstates  have similar coupling to gluons and their mass-scale relation 
 deviates from that for the standard QCD scenario (Eq.~(\ref{invisiblesaxion})).

 This behaviour can be understood as given by the  {\it axionness} of each eigenstate $\beta_i\equiv1/g_i$, and noting that a sum rule is strictly obeyed:
\begin{equation}
\label{eq:sum2}
\frac{1}{g_1} + \frac{1}{g_2} = \beta_1 + \beta_2= \frac{\chi_{\rm QCD}}{m_1^2 f_1^2} + \frac{\chi_{\rm QCD}}{m_2^2 f_2^2} = 1\,.
\end{equation}
This constraint illustrates how the  QCD topological susceptibility is shared among the two axion eigenstates. It also implies that, whenever one of the two eigenstates decouples, the other converges towards the standard QCD axion. An intuitive interpretation of the $\beta_i$-factors can be found 
in App.~\ref{sec:new_physical_derivation}, where it is shown that for a physical axion to decouple from the sum rule it should either have a negligible PQ charge or a negligible projection onto the field (or combination of fields) coupling to $G \widetilde G$,  $\agg$. 

Examples of multi-axion solutions to the strong CP problem 
are illustrated in Fig.~\ref{fig:fa_ma_example}. If a experiment detects a signal at the location indicated by the star, and it  corresponds to a pure QCD axion solution, it is predicted that 
\begin{itemize}
\item No signal should be found inside all the dashed  area --delimitated by a grey line, including in particular the standard -yellow- QCD line.
\item A second axion signal should be found in the undashed area. For the $N=2$ and the toy model above, its location is marked in the figure by an orange diamond.
\end{itemize}
The first point stems from the sum rule in Eq.~(\ref{eq:sum2}), and in fact it will hold whatever the total number of axion eigenstates, see next section.
There, we will generalize that sum rule
to an arbitrary number of axion eigenstates for generic PQ-invariant potentials.

In addition, Fig.~\ref{fig:gN2}  indicates 
 that, for the eigenstate closest the the standard QCD band, the 
point of largest departure 
corresponds to $g_1=g_2=2$ (case not illustrated in Fig.~\ref{fig:fa_ma_example}); this fact will turn out to extend to arbitrary $N=2$ mixing potentials\,\footnote{In the particular example above, the $r$ value which allows maximal departure of the eigenstate closest to the standard QCD band is $r=2$. Using the same simple potential, but for different $\hat f_1$ and $\hat f_2$, that maximal deviation is given by $r=(\hat f_1^2 + \hat f_2^2)/\hat f_1^2$.}.

The results above for the $\{m_i, 1/f_i\}$ QCD band apply directly to experiments which are independent of the axion UV complete models, such as CASPEr-electric~\cite{JacksonKimball:2017elr} or future experiments based on the piezoaxionic effect~\cite{Arvanitaki:2021wjk}. It is easy to prove, though, that the couplings to photons and nucleons get analogously displaced under some simplifying assumptions, see \cref{subsec:coupling_to_photons} for this analysis in the generic $N$ axion scenario.
 \begin{center} 
 \begin{figure}[t]
 \includegraphics[scale=0.263]{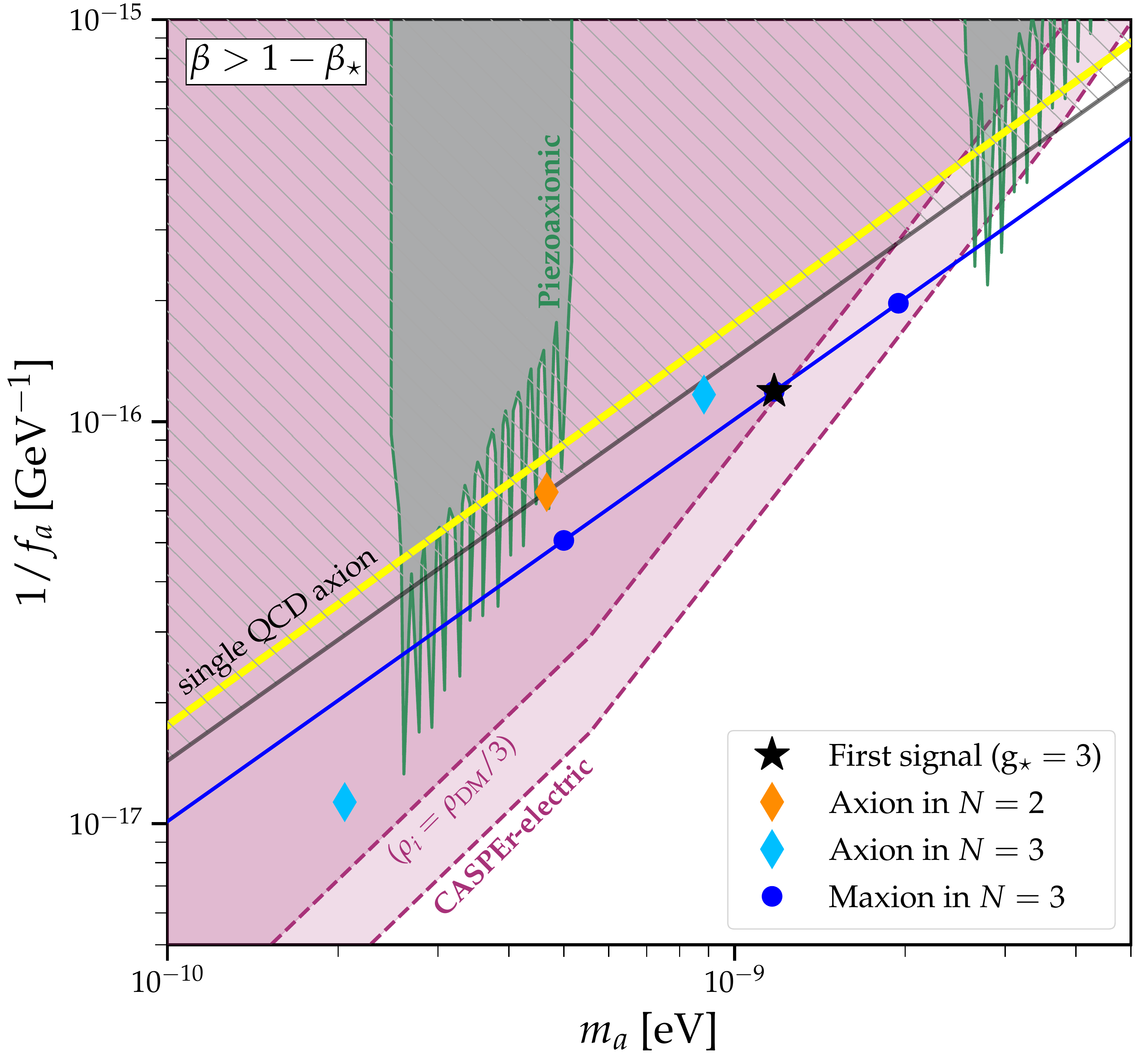}
 \caption{Examples of multi-axion solutions to the strong CP problem discussed in the text. It is assumed that the first experimental ALP signal is observed at the benchmark point represented by a star, with $\beta_\star = 1/3$. The gray dashed region is then excluded if the star describes a true QCD axion (i.e. the PQ symmetry is only broken by QCD). The projected limits for CASPEr-electric (phase III)~\cite{JacksonKimball:2017elr} and piezoaxionic experiments are also represented~\cite{Arvanitaki:2021wjk}. Axion limits taken from Ref.~\cite{AxionLimits}.{\hfill}
 }
  \label{fig:fa_ma_example}
 \end{figure}
 \end{center}
%

\section{Multiple QCD axion fields}
\label{multiple}
It will be shown next that whenever the axion field  (or field combination) that couples to ${G\widetilde G}$ mixes with other $N-1$ scalar singlets via a PQ-invariant potential,  it will lead in all generality to $N$ axion eigenstates, i.e. $N$ distinct massive pseudoscalars, each of them coupling to   ${G\widetilde G}$ in a very specific way.

Let us first formulate the general case of a system of $N$ scalar fields $\{{\hat a}_k\}_{k=1}^N$ which couple to the QCD topological term through one arbitrary linear combination, and are in addition subject to an extra scalar potential $V_B({\hat a}_k)$ which mixes them, 
\begin{align}
\label{Lgeneral}
\mathcal{L}=\frac{\alpha_s}{8 \pi}\left(\sum_{k=1}^{ N} \frac{{\hat a}_k}{{\hat f}_k}-\bar{\theta}\right) G\widetilde G  -V_B({\hat a}_1,\,{\hat a}_2,\dots, {\hat a}_N)\,,
\end{align}
where, without loss of generality, the decay constants ${\hat f}_k$ have been redefined to absorb any coefficient in the coupling to gluons. Without the extra potential, the system would have exhibited $N$ independent global U(1) symmetries at the classical level, with one combination of the $N$ GBs becoming the standard massive QCD axion. The mixing potential reduces the symmetry. 
To be precise, note that neither the number of fields that couple to $G \widetilde G$ nor  the dimension of the mixing potential $V_B$ needs to coincide with the number of scalars in Nature $N$ which mix, but the results below are independent of it, as shown  in the basis of Eq.~(\ref{Lgeneral-rotated}).

The condition 
to solve the strong CP problem in the presence of the potential is simply that the latter must exhibit a $U(1)_{PQ}$ symmetry, i.e. solely broken by QCD.
 Imposing this \emph{true PQ symmetry} ensures that the minimum of the complete potential corresponds to the CP conserving point, $\bar \theta_{\text{eff}}=0$, and it will have important consequences on the properties of the different axions. From now on we will assume that all axion fields are already redefined and expanded around their minimum and thus we will drop the $\bar \theta$-term.

It is instructive to consider the complete scalar mass matrix which stems from Eq.~(\ref{Lgeneral}). At energies below the QCD confinement scale, the gluonic coupling induces a non-zero potential\footnote{The leading order QCD axion potential~\cite{DiVecchia:1980yfw,Leutwyler:1992yt,diCortona:2015ldu} reads $V({\hat a}_k)=-m_{\pi}^{2} f_{\pi}^{2}  \sqrt{1-\beta \sin ^{2}\left(\frac{1}{2} \sum\frac{{\hat a}_k}{{\hat f}_k}-\bar{\theta}\right)}$, where ${\beta\equiv\frac{4m_um_d}{(m_u+m_d)^2}}$.} which corresponds to the following mass term,
\begin{align}
 \frac{\alpha_s}{8 \pi}\sum_{k=1}^N \frac{{\hat a}_k}{{\hat f}_k} \, G\widetilde G  \quad 
\longrightarrow \quad \frac{1}{2}\chi_{\text{QCD}}\left(\sum_{k=1}^N \frac{{\hat a}_k}{{\hat f}_k}\right)^2\,,
\end{align}
where the QCD topological susceptibility $\chi_{\text{QCD}}$ was defined on the right-hand side (RHS) of Eq.~(\ref{invisiblesaxion}).  
The complete mass matrix $\hat{\mathbf{M}}$ can then be  defined as
\begin{align}
\mathcal{L}\supset -\frac{1}{2} {\hat a}_k \hat{\mathbf{M}}_{k l}^2  {\hat a}_l \quad \text{ with } \quad \hat{\Mbf}^2= \hat{\mathbf{M}}_A^2+\hat{\Mbf}_B^2 \,,
\label{Eq:MAhatMBhat}
\end{align}
where 
$$\big(\hat{\Mbf}_A^2\big)_{k l}\equiv  \frac{\chiq} {{{\hat f}_k{\hat f}_l}}\,$$ has rank $1$ and encodes the QCD contribution to the masses of the axion eigenstates, while ${\big(\hat{\Mbf}_B^2\big)_{k l}\equiv \partial^2 V_B/\partial a_k\partial a_l}$ stems from the extra potential.

If the system has a PQ symmetry, the determinant of the complete mass matrix $\hat{\Mbf}$ needs to vanish in the limit where the QCD topological susceptibility goes to zero, which in turn implies that the determinant of  $\hat{\Mbf}_B$ vanishes\footnote{\label{PQMixingCondition}In order for the result to apply both ways, one needs to add on the RHS the condition that the eigenvector of $\hat \Mbf_B^2$ with vanishing eigenvalue, $a_{\rm PQ}$ (see App.~\ref{sec:new_physical_derivation}),
has a non-zero projection to the linear combination that couples to gluons,
i.e. $\langle  a_{\rm PQ}|\agg\rangle\neq 0$.
},
\begin{align}
\exists\,  U(1)_{PQ} \Rightarrow \lim_{\chiq\rightarrow0} \det \hat{\Mbf}^2=0 \Rightarrow \det \hat{\Mbf}_B=0\,.
\label{Eq:PQdetMB=0}
\end{align}

\textbf{A rotated basis.} The problem can be equivalently formulated rotating to a field basis in which  the combination of fields coupled to $G\widetilde G$ is redefined as a single field $ {\hat a}_{G\widetilde G}$,
\begin{equation}
\label{aGdual}
\frac{{a}_{G\widetilde G}}{F} \equiv \sum_{k=1}^N \frac{{\hat a}_k}{{\hat f}_k}\,,\qquad   \textrm{with }\quad
\frac{1}{F^2}= \sum_{k=1}^N \frac{1}{{\hat f}_k^2}\,, 
\end{equation}
and as a consequence ${\hat a}_{G\widetilde G}$ mixes with the remaining   $N-1$  fields via the (rotated) external potential $V^{\textrm{R}}_B$,
\begin{align}
\label{Lgeneral-rotated}
\mathcal{L}=\frac{\alpha_s}{8 \pi}\frac{{a}_{G\widetilde G}}{F} G\widetilde G  - V^{\textrm{R}}_B(\underbrace{{a}_{G\widetilde G},\,\dots}_{N})\,.
\end{align}
The relation of the  scale $F$ to the $\{\hat f_k\}$ set in Eq.~(\ref{aGdual}) follows from  field normalization.
In this perspective, the field coupling to ${G\widetilde G}$ is allowed to mix with other  singlet scalars in Nature through a PQ-invariant potential. This  is the generalization of the rotation performed earlier for $N=2$ between  Eqs.~(\ref{eq:L2}) and Eqs.~(\ref{eq:L2-otro}). The rotated basis will  be preferred further below, as it renders particularly straightforward  mathematical demonstrations.

\subsection*{Peccei-Quinn condition}
 
In the rotated basis, the square mass matrix ${\mathbf{M}^2\equiv \mathbf{R}\,\hat{\Mbf}^2 \mathbf{R}^T}$ can again be decomposed in two parts, where the part proportional to  $\chiq$, $\Mbf_A^2$, is now diagonal and has only one non-zero element corresponding to $ { a}_{G\widetilde G}$,
\begin{align}
\mathbf{M}^2&  
\label{Eq:MassMatricesPreferredbasis-first}
= \mathbf{M}_A^2+\mathbf{M}_B^2 
= \left(
  \begin{array}{cc}
   b_{11}  &  \mathbf{X}^T \\
   \mathbf{X} &  \mathbf{M}^2_1  \\
  \end{array}
  \right)\\ 
  &=
\frac{\chiq}{F^2}\,
 \left(
  \begin{array}{cc}
   1  &  0 \\
   0 &  \mathbf{0}  \\
  \end{array}
  \right)
  + 
  \left(
  \begin{array}{cc}
   b_{11}-\frac{\chiq}{F^2}  &  \mathbf{X}^T \\
   \mathbf{X} &  \mathbf{M}^2_1  \\
  \end{array}
  \right)\,,
  \label{Eq:MassMatricesPreferredbasis}
\end{align}
in which both $\mathbf{0}$ and the minor\footnote{By minor matrix $\mathbf{M}_1^2$ we mean the  $(N-1)\times(N-1)$ submatrix resulting from deleting the first row and column.} $\mathbf{M}_1^2$ are $(N-1)\times(N-1)$ matrices, $\mathbf{X}$ is a $(N-1)$-dimensional column vector  
and $b_{11}$ is a scalar.
The PQ-invariance condition in Eq.~(\ref{Eq:PQdetMB=0})  reads now,
\begin{align}
\det {\hat \Mbf}_B=\det (\mathbf{R}\,\mathbf{M}_B^2 \mathbf{R}^T)=0\,,
\label{Eq:PQdetMBrotated=0}
\end{align}
and applying the Schur complement it follows that 
\begin{align}
\det \Mbf_1^2 \left(b_{11}- \frac{\chiq}{F^2}  - \mathbf{X}^T \Mbf_1^{-2}\mathbf{X}\right)= 0\,.
\label{Eq:Schur complement}
\end{align}
Assuming that all axions become massive\footnote{Our result can also be applied if some axions were to be massless. If $m$ out of the $N$ eigenstates are  massless, one can simply block diagonalize those scalars first (which in fact decouple from the QCD sector), and then apply the result to the $(N-m)\times (N-m)$ dimensional matrix. }, i.e. $\det \Mbf_1^2\neq 0$, the second factor in \cref{Eq:Schur complement} must vanish and applying again the Schur complement to the complete matrix in the rotated basis, $\Mbf^2$ in Eq.~(\ref{Eq:MassMatricesPreferredbasis-first}), we obtain in all generality that the necessary  condition for a scalar potential  to be PQ invariant is  
\begin{align}
\frac{\det \Mbf^2}{\det \Mbf_1^2}= \left(b_{11}  - \mathbf{X}^T \Mbf_1^{-2}\mathbf{X}\right)=\frac{\chiq}{F^2}\,,
\label{Eq:PQrelation-chi}
\end{align}
i.e. 
\begin{align}
\label{Eq:PQrelation}
\frac{\det \Mbf^2}{\det \Mbf_1^2}=\frac{f_\pi^2 m_\pi^2}{{F}^2}  \frac{m_u m_d}{\left(m_u+m_d\right)^2} \,,
\end{align}
plus the mixing condition stated in \cref{PQMixingCondition}. 
For a given mass eigenstate $|a_i\rangle$ with eigenvalue  $m_i$, let    $f_i$ denote the physical strength of its coupling to gluons,
 \begin{align}
\mathcal{L}\supset\frac{\alpha_s}{8 \pi} \frac{a_i}{f_{i}} G\widetilde G\,.
\end{align}
 Denoting  by $v_{ij}$  the elements of the rotation matrix that diagonalizes $\Mbf^2$ (which define the eigenvectors of the system),  the projection of $|a_i\rangle$ onto the unit vector which couples to gluons reads 
\begin{align}
v_{i1} = \langle \agg|a_i\rangle \,.
\end{align}
 It then follows that $m_i$ can then be expressed as 
\begin{align}
\label{mass-eigenvalues}
\quad m_{i}^2
=\left|v_{i1}\right|^2 \frac{\chiq}{F^2}
+\langle a_i|\Mbf_B^2|a_i\rangle\,,  
\end{align}
while
\begin{align}
\label{physical_fi}
\frac{1}{f_i}=\frac{1}{F}\times  |v_{i1}|\,  \implies  \frac{1}{F^2}= \sum_i  \frac{1}{f_i^2}\,,
\end{align}
that is, the physical coupling to gluons of a given axion eigenstate can be seen as the projection of the  coupling of  $\agg$  into that eigenstate.

\subsection{Distance from the standard axion band} 
\label{sec:qcd_axion_likeness}

The physical, measurable, quantities are $m_i$ and $f_i$. 
How much can their value  depart from the standard $\left\{m_a, 1/f_a\right\}$ QCD band,  defined by Eq.~(\ref{invisiblesaxion})?

The factor of $g_i$ defined in \cref{Eq:gfactors} precisely describes how much the mass-scale relation for an eigenstate $a_i$ deviates from that for the single QCD axion. In light of Eq.~(\ref{physical_fi})  the $g_i$ factors can be expressed in the preferred basis as 
\begin{align}
\label{gi}
g_i=\frac{m_{i}^2 \, F^2}{\left|v_{i1}\right|^2\chiq}\,, 
\end{align}
which in turn implies  in terms of  $\beta_i=1/g_i$: 
\begin{align}
\quad m_{i}^2
=\beta_i \, m_{i}^2
+\langle a_i|\Mbf_B^2|a_i\rangle \,,
\label{Eq:PQNessFraction}
\end{align}
which shows that $\beta_i$ is the fraction of a given mass eigenvalue $m_i$ due to QCD, its ``\emph{axionness}''.  
\paragraph*{\bf{A sum rule.}}

 It is fruitful to apply now the eigenvalue-eigenvector identity~\cite{JacobiDeBQ,Denton:2019pka} (see \cref{Eq:EigenvalueEigenvectorIdentity}). In one of its versions it reads
 \begin{equation}
\frac{\text{det} \left(\lambda \mathbb{I}_{N-1} - M_j\right)}{\text{det} \left(\lambda \mathbb{I}_{N} - A\right)} = \sum_{i=1}^N \frac{|v_{ij}|^2}{\lambda - \lambda_i}\,,
\end{equation}
 where  $A$  denotes any Hermitian matrix with eigenvalues $\{\lambda_i(A)\}$ and associated eigenvectors $v_{ij}$, and where $M_j$ denotes the minor matrix formed by removing the $j^{\text {th }}$ row and column of $A$. Applying this identity to the mass matrix in the preferred basis in \cref{Eq:MassMatricesPreferredbasis}, $A=\Mbf^2$, it implies for $j=1$ and $\lambda=0$ that 
\begin{align}
\frac{\det \Mbf_1^2}{\det \Mbf^2}= 
\sum_{i=1}^N \frac{ \left| v_{i1} \right|^2}{m_{i}^2}
=\frac{F^2}{\chiq}\sum_{i=1}^N \frac{1}{g_i}.
\end{align}
By comparison with the requirement that the potential has a \emph{true PQ symmetry},  \cref{Eq:PQrelation}, the following constraint on the possible values of the $g_i$ factors results: 
\begin{align}
\Aboxed{\exists\, \,  U(1)_{PQ} \implies \sum_{i=1}^{N} \frac{1}{g_i} =1\,,}
\label{eq:gi-1constraint PQ}
\end{align}
or equivalently
\begin{equation}
\label{eq:sum-rule}
  \sum_{i=1}^{N} \beta_i =1\,. 
  \end{equation}
  An alternative derivation of this QCD axion sum rule can be found in App.~\ref{sec:new_physical_derivation}, which provides an intuitive understanding of  \emph{axionness}}.

This sum rule links the multiple QCD axion signals and allows a direct comprehension of the decoupling limits in which only one single QCD axion is reachable.~{In order to obtain a different phenomenology with respect to this standard case, at least one scale in the mixing potential is required to be of the order of the QCD induced mass scale, a fact we demonstrate in App.~\ref{app:Vscale}.}

Furthermore, a number of properties of the multiple axion eigenstates follow from our sum rule, as we will rigorously prove next.

\begin{enumerate}
  \item All $g_i$-factors are larger or equal than one, 
  $$g_i\geq 1\, $$   
  (i.e. $\beta_i\leq 1$), as a consequence of the positivity of the extra potential. i.e. of $\Mbf_B^2$ being positive semi-definite. For a \emph{true PQ symmetry,} this implies that  the  QCD axion line  in  the $\left\{m_i, 1/f_i \right\}$  parameter space of any of {\it these multiple axions can only be displaced to the right} of that of the single axion case, by a factor $\sqrt{g_i}$.

  \emph{Proof}. 
 Combining the expression for the mass eigenvalues in the rotated basis, Eq.~(\ref{mass-eigenvalues}), with Eq.~(\ref{gi}), it results that 
\begin{align}
g_i=1+\frac{F^2}{\chiq}\frac{\langle a_i|\Mbf_B^2|a_i\rangle }{\left|\langle a_i|\agg\rangle\right|^2}\geq1
\end{align}
since $\Mbf_B^2$ is semi-positive definite\footnote{The mass matrix of a system of axions is defined as the second derivative in the minimum of the potential and thus it is semi-positive definite. However, if one splits the mass matrix in two contributions, nothing guarantees that both matrices are still semi-positive definite. We argue, nevertheless, that this is indeed the case for our matrices. Due to the PQ symmetry, $\Mbf_B^2$ has vanishing determinant and thus there are more axions than constraints in the extra potential in order to find the minimum. 
}. It follows that $\beta_i \le 1\,, \forall i$. 

\item If a given axion $a_i$  has a factor $g_i\ne 1$, the sum rule sets an upper limit on the  $g$-factor of any other axion,
  \begin{align}
  g_j\geq \frac{1}{1-1/g_i}\,\quad \forall \, j\neq i\,,
  \end{align}
  or, equivalently
   \begin{align}
  \beta_j\leq 1-\beta_i\,\quad \forall \, j\neq i\,.
  \end{align}
   \emph{Proof}.  This follows directly from the inequality
   \begin{align}
 1-\frac{1}{g_i}=\sum_{j\ne i}^{N} \frac{1}{g_j}\geq \frac{1}{g_j}\,.
\end{align}
 This bound has interesting experimental consequences as we will discuss in \cref{sec:experimental_consequences}. 
 \item As a particular case of the previous property, if one axion is detected with $g_i=1$, then no other axion eigenstate can couple to gluons and thus no other axion signal is to be expected. Conversely, it is enough to detect one axion signal with $g_i\ne1$ to imply that at least another axion signal awaits discovery. 
   \item {\bf Maximal distance of the axion closest to standard: Maxions. 
   }
   The maximum value that the smallest of all $g_i$'s can take is $N$, the number of physical axions,  and it  corresponds to the case where all of them are equal, 
  \begin{align}
  \label{maxion-definition}
      \qquad \max
       \left\{\min_i\{g_i\}\right\}=N \quad \implies g_i=N,\,\,\forall \,i \,.
    \end{align} 
    This follows directly from the sum-rule in Eq.~(\ref{eq:gi-1constraint PQ}), that shows that if any $g_i$ deviates from the point where all are equal, another one has to deviate contrarily following a see-saw pattern.  In the scenario in Eq.~(\ref{maxion-definition}), all the $N$ axion signals will appear aligned over just one QCD line parallel to that for the single QCD axion, and displaced to its right by a factor $\sqrt{N}$. 
    We will denominate this type of solutions as    \emph{QCD maxions} (maximally deviated QCD axions). 
    
    For maxions, the coefficients  $c^\Mbf_k$ of the characteristic polynomial of $\Mbf^2$,
   \begin{align}
 \mbox{\large $p$}_{{\,\Mbf^2}}(\lambda)&\equiv \sum_{k=0}^{N} c^\Mbf_k \lambda^k\,,
\end{align}
and the analogous ones for the minor $\Mbf_1^2$,  $c^{\Mbf_1}_k$, 
fulfill the relation
\begin{align}
\label{maxion-coeffs}
    c^\Mbf_k=- N\, \frac{\chiq}{F^2 (N-k)}\, c^{\Mbf_1}_k\,.
    \end{align}
    These coefficients can be expressed in all generality in terms of the complete exponential Bell polynomials, 
    \begin{multline}
    \qquad c_{n-k}^\mathbf{A}=\frac{(-1)^{n-k}}{k !} \mathcal{B}_k\big(\operatorname{tr} A,-1 ! \operatorname{tr} A^2, 2 ! \operatorname{tr} A^3,\\ 
     \ldots,(-1)^{k-1}(k-1) ! \operatorname{tr} A^k\big) \nn
    \end{multline}
    and then  the relation in Eq.~(\ref{maxion-coeffs})    reads
    \begin{align}
    \label{maxion-coeffs-Bell}
    \mathcal{B}_{N-k}^{\Mbf^2}=N \,\frac{\chiq}{{F}^2}\,\mathcal{B}_{N-k-1}^{\Mbf_1^2}\,.
    \end{align}
This constitutes a set of $N$ \emph{QCD maxion  conditions} necessary to solve the strong CP problem via maxions. For the particular case  in which all  decay constants  in Eq.~(\ref{Lgeneral}) are equal, ${\hat f}_k={\hat f}$, it follows that $F=\hat f/\sqrt{N}$ and these relations simplify to 
\begin{align}
    \label{maxion-coeffs-Bell}
    \mathcal{B}_{N-k}^{\Mbf^2}= N^2\,\frac{\chiq}{{\hat f}^2}\,  \mathcal{B}^{\Mbf_1^2}_{N-1-k}\,.
    \end{align}
    A particular relation of interest is that for ${k=N-1}$, which implies the following constraint the sum of all  square maxion masses: 
\begin{equation}  
\label{eq:mTr}
\mathcal{B}_{1}^{\Mbf^2}= \text{ tr}\,{\mathbf{M}^2}=\sum_{i=1}^N m_i^2=N \frac {\chiq}{ F^2}\,.
\end{equation}
Furthermore, for $k=0$ the maxion condition boils down to the PQ constraint on the determinants of a generic potential, \cref{Eq:PQrelation}.
 The proof of the relations in Eq.~(\ref{maxion-coeffs-Bell}) is somewhat lengthy, and we have thus deferred it to App.~\ref{sec:qcd_maxion_condition}.

Interestingly, the maxion conditions are automatically fulfilled by Laguerre polynomials, as explained in \cref{sec:applications}. Mixing matrices in clockwork scenarios~\cite{Kim:2004rp,Choi:2014rja,Kaplan:2015fuy,Giudice:2016yja}, on the contrary, do not in general satisfy these maxion conditions, as shown in App.~\ref{app:clockwork}.
\end{enumerate}

An illustration of the results in this section is provided in  Fig.~\ref{fig:fa_ma_example}, for  generic $N=2$ and $N=3$, patterns  as well as a  $N=3$ maxion solution, see discussion in ~\cref{sec:experimental_consequences}.

\subsection{Extension to the coupling to photons} 
\label{subsec:coupling_to_photons}
%
\begin{figure}[t]
\centering
\includegraphics[scale=0.25]{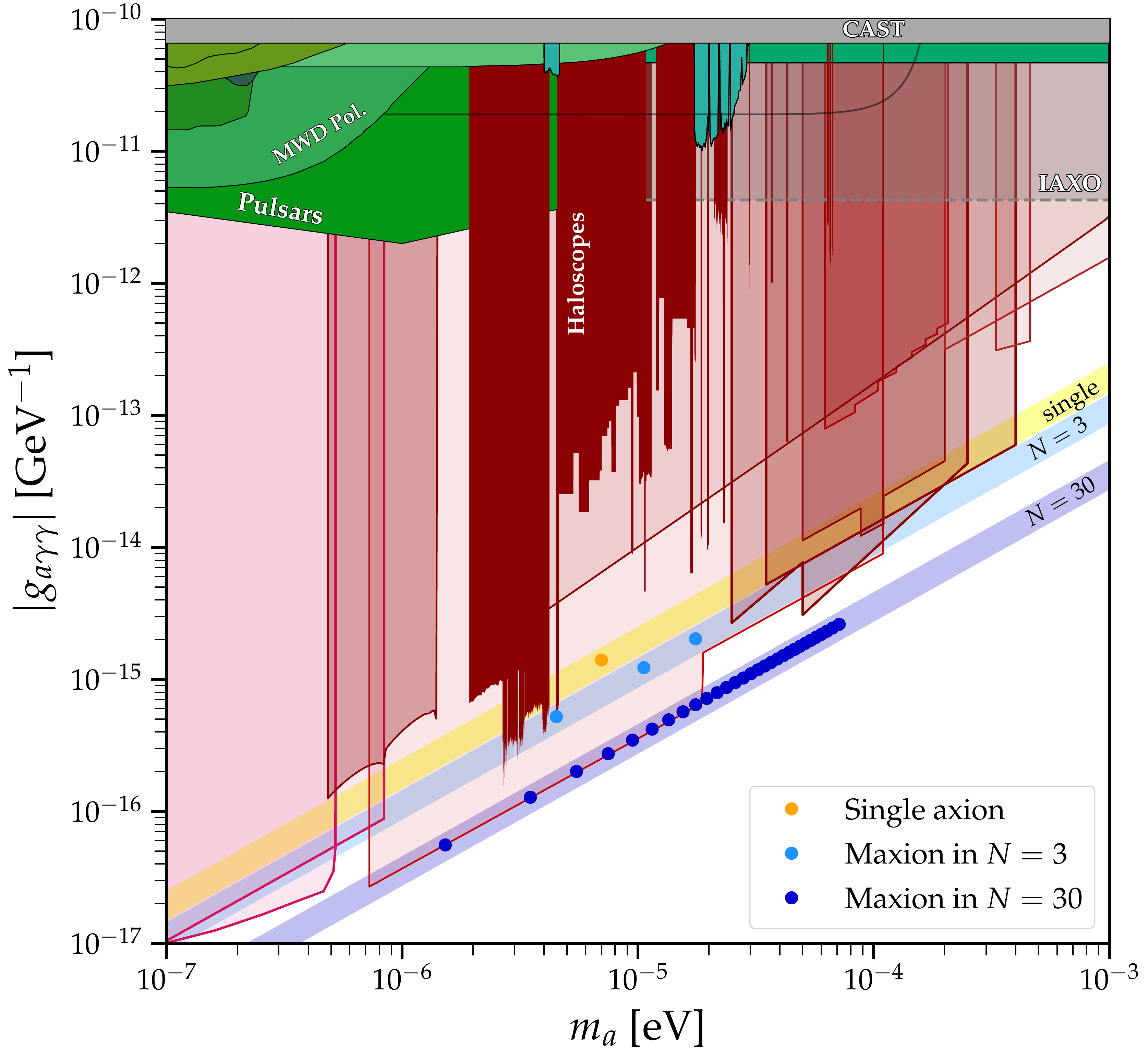}
\caption{Predictions for the photon coupling of multiple QCD Laguerre maxions; see the text for details. The width of each colored band spans $E/\mathcal{N}\in [2/3,8/3]$. The colored regions represent the parameter space currently excluded, while the light red regions show the projected limits for several haloscope experiments. Axion limits adapted from Ref.~\cite{AxionLimits}. {\hfill}
}
\label{fig:fa_photons}
\end{figure}
%

It is not completely straightforward to extend the constraints studied above for the axion-gluon coupling to the interactions to photons. The reason is that the axion-photon coupling has a component that depends  on the UV axion model through the ratio $E/\N$, where $E$ denotes the model-dependent  electromagnetic anomaly and $\N$  the color anomaly.  

A simple case is that in which  $E/\N$ is assumed to be equal for all axion fields  $\hat a_k$ in Eq.~(\ref{Lgeneral}), ${{E_k}/{\N_k}={E}/{\N}}$, $\forall k$. The model-dependent axion-photon interactions read then
\begin{align}
\label{Lelectric}
\delta \L= \frac{1}{4}\sum_{k=1}^{ N}g^0_{{\hat a}_k\gamma\gamma}\, {{\hat a}_k} F\widetilde F \,\equiv \frac{\alpha_{em}}{8 \pi}\sum_{k=1}^{ N}\frac{E}{\N}\frac{{\hat a}_k}{{\hat f}_k} F\widetilde F \,,
\end{align}
which in terms of the preferred rotated basis  becomes
\begin{align}
\label{Lelectric}
\delta \L= \, \frac{\alpha_{em}}{8 \pi} \frac{E}{\N}\frac{a_{G \tilde G}}{F} F\widetilde F \,,
\end{align}
i.e. the axion-photon coupling is aligned with the gluonic one, see Eq.~(\ref{Lgeneral-rotated}).   This already suggests that the fate of the standard band for the QCD axion-photon coupling will follow a very similar pattern to that 
determined for the coupling to gluons. 
 The physical coupling to photons will receive as usual a  model-independent QCD contribution stemming from the axion-$\eta'$-pion mixing. This component can be easily computed by performing an axion-dependent rotation on the quark fields \cite{Georgi:1986df,diCortona:2015ldu},
\begin{align}
q=\left(\begin{array}{l}
u \\
d
\end{array}\right) \rightarrow e^{i \gamma_5 {a_{G\tilde G}}/({2 F}) Q_a}\left(\begin{array}{c}
u \\
d
\end{array}\right), 
\end{align}
where the matrix $Q_a$ is chosen so that the  tree-level  axion-pion mixing  is avoided,  ${Q_a = {M_q^{-1}}/{\text{tr}\,(M_q^{-1})}}$, where  ${M_q = \text{diag}(m_u,\,m_d)}$ denotes the quark mass matrix. In other words, it is chosen so that the  mass matrix for $\{a_{G\tilde G},\pi^0\}$ is automatically diagonal. This anomalous rotation modifies the coupling to photons as customary with a shift on $E/\N$ of $-6 \operatorname{tr}\left(Q_a Q^2\right)$ -where $Q$ is the electric quark charge matrix- leading to~\cite{diCortona:2015ldu},
\begin{align}
\frac{E}{\N}\longrightarrow 
\frac{E}{\N}-\frac{2}{3} \frac{4 m_d+m_u}{m_d+m_u}
\xrightarrow[]{\text{NLO}}\frac{E}{\N}-1.92\,,
\end{align}
where in the last step next-to-leading corrections in the chiral expansion have been incorporated. Finally, in terms of the mass eigenvalues $m_i$ and physical scale $f_i$ of the eigenstates $a_i$, the axion-photon interactions, for this universal $E/\N$ case, are given by
\begin{align}
\L\supset\frac{\alpha_{e m}}{2 \pi }  \left[\frac{E}{\N}-1.92\right] \sum_i\frac{a_i}{f_i}\,
F\widetilde F\,.
\end{align}
This implies that the single QCD axion band in the coupling to photons will be shifted by the same factor of $g_i$ that has been determined for the gluon-axion coupling in the previous section, 
\begin{align}
\frac{m_i^2}{g_{a_i\gamma\gamma }^2}=\frac{m_a^2}{g_{a\gamma\gamma }^2}\Bigg|_{\text{single QCD axion}}\times g_i\,.
\end{align}
 A sum rule for the coupling to photon of the $N$ axion excitations follows (analogous to that for the PQ condition in Eq.~(\ref{eq:gi-1constraint PQ})),
\begin{align}
\frac{(2 \pi)^2 \chiq}{\alpha_{e m}^2}\left[\frac{E}{\N}-1.92\right]^{-2}\,\sum_{i=1}^N\frac{g_{a_i\gamma\gamma}^2}{m_i^2}=1\,.
\label{Eq:COnstraintPhotons}
\end{align}
Two plausible scenarios in which $E/\N$ would be universal for all axions -and thus the results above hold directly- are: i) the UV models in which no exotic fermions are electromagnetically charged i.e. $E/\N=0$, and ii) when the SM is embedded in 
a Grand Unified Theory (GUT), which can fix $E/\N=8/3$~\cite{Srednicki:1985xd,Agrawal:2022lsp}.  
In fact, the authors of Ref.~\cite{Agrawal:2022lsp} already noted that, in the presence of multiple axions in a GUT theory, their photon-couplings were always located to the right of the standard axion-photon band.  
They did not explore the fundamental axion-gluon coupling, though, neither determined the maximal distances nor the general proofs and sum-rules obtained here.

The more general case in which  $E/\N$ differs for each of the axion fields can be worked out from the results above.

\section{Maxion masses}
\label{sec:applications}
Even though the maxion conditions require the presence of mass scales in the extra potential which are of the same order as the QCD-induced mass, and restrict the shape of the matrices, there are still infinitely many maxion matrices. Indeed, the class of matrices that verify these QCD Maxion conditions generate a $m$-parameter family of matrices with $m=N(N+1)/2$. 
    Fixing all $\hat{f}_i = \hat{f}$, the number of families reduces to $m=N(N-1)/2 +1$. Consequently, in the latter case all $N=2\,(3)$ maxion potentials are functions of $1\,(3)$ free parameter(s), plus the overall scale.

In a scenario with 2 axion fields produced  with  the same decay constant ($\hat f_1= \hat f_2= \hat f$), the only family of maxions is characterized by the following mass matrix:
\begin{equation}
\label{eq:maxionN2}
\hat{\mathbf{M}}^2 = \frac{\chi_{\rm QCD}}{\hat{f}^2}\begin{pmatrix}
3 - r & 1+\sqrt{r(2-r)} \\
1+\sqrt{r(2-r)}  & 1+r
\end{pmatrix}\,,
\end{equation}
where $r\in [0,1) \cup (1,2]$. For $r=2$, the toy example presented in Sec.~\ref{sec:N2} is recovered, while $r=1/5$ reproduces the example in App.~\ref{App:UV}. 

The eigenvalues obtained from this matrix lead to ${\Delta \equiv |m_1^2 - m_2^2|/(m_1^2+m_2^2)}$ ranging from $\sim 0.7-1$. In the limiting case  in which $\Delta = 1$, corresponding to $r=1$, the system is not PQ-invariant, as the massless eigenstate has no mixing with $a_{G\widetilde{G}}$; see \cref{PQMixingCondition}. Nonetheless, values of $\Delta = 1+\epsilon$ (with $\epsilon \ll 1$) are possible, corresponding to an almost massless maxion and another one with 
$m_2^2 \approx  2\chi_{\rm QCD}/ F^2$. 
This shows that \textit{highly hierarchical masses are possible}. Assuming a different hierarchy for $\hat f_1$ and $\hat f_2$, values of $\Delta \sim 0$ can be attained in addition.

\begin{figure}[t]
\includegraphics[scale=0.4]{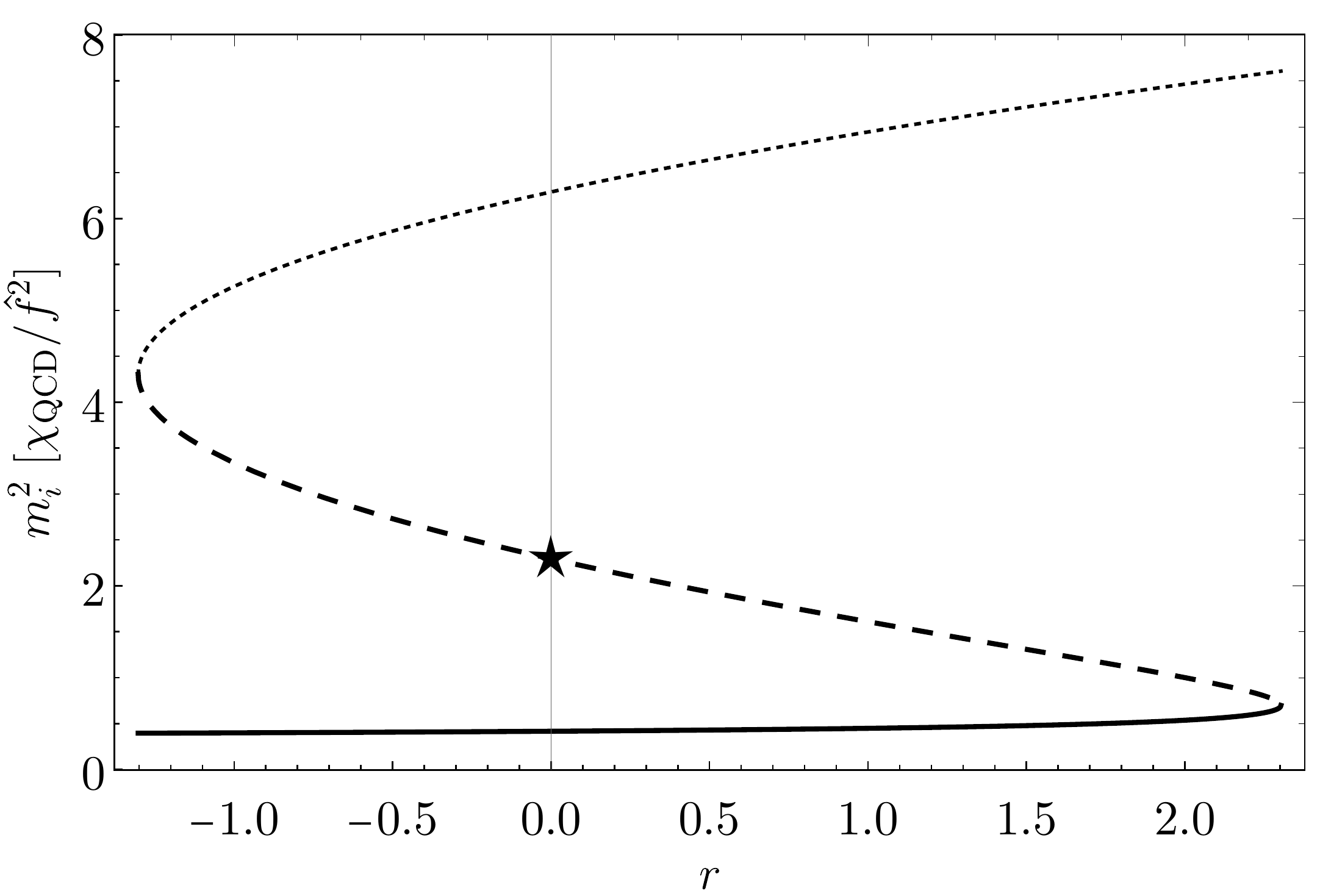}
\caption{Spread in maxion masses, for $N=3$ and $\hat{f}_i = \hat{f}$, for the particular type of scalar potentials in \cref{eq:maxionN3}, characterized by a single parameter $r$. The solid, dashed and dotted lines show, respectively, the mass of the lightest, intermediate and heaviest maxion. The star is the illustrative eigenvalue represented in Fig.~\ref{fig:fa_ma_example}. {\hfill}
}
\label{fig:mN3}
\end{figure}

In scenarios with 3 axions, the maxion families are instead characterized by 3 free parameters, assuming equal decay constants for all fields. Identifying $\hat{a}_1$ with the PQ combination, an  illustrative example is given by the mass matrix 
\begin{equation}
\label{eq:maxionN3}
\hat{\mathbf{M}}^2 = \frac{\chi_{\rm QCD}}{\hat{f}^2}\begin{pmatrix}
1 & 1 &1 \\
1 & 4 - \sqrt{3 + r - r^2} & 1+ r \\
1 & 1+r & 4 + \sqrt{3 + r -r^2}
\end{pmatrix}\,,
\end{equation}
with $r\in \left[\frac{1}{2} (1-\sqrt{13}), \frac{1}{2} (1+\sqrt{13})\right]$. The eigenvalues sourced by this mass matrix are shown in Fig.~\ref{fig:mN3}. 
If it was furthermore assumed that the extra source of interactions $V_B$ is diagonal in the scalar fields, it would follow that:
\begin{equation}
\label{eq:MN3}
\hat{\mathbf{M}}^2_{p=0} = \frac{\chi_{\rm QCD}}{\hat{f}^2}\begin{pmatrix}
1 & 1 &1 \\
1 & 4-\sqrt{3}& 1\\
1 & 1 & 4+\sqrt{3}
\end{pmatrix}\,.\end{equation}
In this particular example, the entries of $\mathbf{M}^2$ in the preferred basis turn out to be simply the roots of Laguerre polynomials, as explained next.

%
%

{\bf Laguerre maxions.} If the matrix $\hat\Mbf_B^2$ in \cref{Eq:MAhatMBhat} is diagonal and all decay constants are equal $\hat f_k=\hat f$, one can show that the maxion conditions are automatically fulfilled if the characteristic
polynomial of the mass matrix and its minor --in the preferred rotated basis--  correspond to the Laguerre polynomials $\pM(\lambda)=L_N(\lambda)$ and $\pMone(\lambda)=L_{N-1}(\lambda)$. We call the solutions produced in this scenario \emph{Laguerre maxions}. Indeed, one can check that for the zeroes $\lambda_i$ of the $N$-th order Laguerre polynomial $L_N(\lambda)$, the following identity holds\footnote{This can be shown using the expression of $g_i$ in \cref{Eq:gfactorDerivative} and the property $\lambda \, d L_N(\lambda)/d\lambda=N L_N(\lambda)-N L_{N-1}(\lambda)$. 
 },
\begin{align}
\qquad g_i\equiv \lim _{\lambda\rightarrow\lambda_i} \frac{ \lambda L_N(\lambda)}{(\lambda-\lambda_i)L_{N-1}(\lambda)}=N
\end{align}
ensuring that all physical axions depart from the standard QCD axion line by a factor of $\sqrt{N}$. The Laguerre Maxion scenario is particularly predictive since the spread in the values of the QCD axion masses will be fixed and determined by the zeroes of the Laguerre polynomials (see example the cases of $N=3, 30$ in \cref{fig:mN3}).

\vspace{1cm}

\section{Discussion and outlook}
\label{sec:experimental_consequences}

The main experimental impact of our results is that multiple axion signals --instead of a single one-- are expected from the generic 
 QCD axion solution to the strong CP problem. The customary solution with a single axion QCD corresponds instead to a certain limit of the general parameter space.  These multiple solutions  can only be displaced towards the right of the customary QCD axion band. 

 In case of a positive  ALP signal to the right of that band (or even within the single QCD band given its error bars), experiments are encouraged to look for similar signals within their frequency range. In addition, signals widely separated and thus accessed by different experiments are possible. This establishes a beautiful synergy between experiments sensitive to vastly different masses and scales, as the complete reconstruction of the solution to the strong CP problem may require this complementary search.

An illustration of this experimental impact on the $\{m_a, 1/f_a\}$ parameter space of axion-gluon couplings  (which are directly tackled by axion-nEDM experiments) is provided in  \cref{fig:fa_ma_example}. Specific  $N=2$ and $N=3$  QCD axion solutions are depicted, corresponding to  mixing matrices in \cref{eq:MN2,eq:MN3} and a generic 3-axion scenario. 
A star indicates the putative location of the first  signal detected in the axion/ALP parameter space. For it to be linked to a pure axion QCD  solution
 of the strong CP problem, it follows that:
\begin{itemize} 
\item No other axion signal can be found anywhere in the greyed area (which includes the single QCD axion band). This holds independently of  the total number of axion eigenstates $N$. 
\item For $N=2$ only a second signal should appear exactly in the grey line, so as to saturate the sum rule; an example is indicated by an orange diamond.
\item For $N=3$,  two other signals await discovery in the undashed area, see e.g. the clear blue diamonds.
\item An $N=3$ maxion solution is also possible in the example chosen,  signaled by the dark blue points, as the distance of the star signal to the single QCD axion band is $\sqrt{3}$. 
\end{itemize}

We have shown that the same pattern holds for arbitrary $N$, for which we have determined in all generality the condition for an arbitrary scalar potential to be PQ invariant. Furthermore, a precise sum rule for the QCD axion has been found, together with other exact results.

For the axion-photon couplings, we have illustrated the analogous multiple signals and displacements in Fig.~\ref{fig:fa_photons}, assuming universal $E/\N$ couplings and focusing on maxion signals: both a $N=3$ and a $N=30$ maxion multiplet are shown, for the mixing provided by the Laguerre matrices described in~\cref{sec:applications}.~The displacements of axion couplings to elementary fermions will be subject to model-dependences analogous to those of axion-photon couplings, to be further explored elsewhere~\cite{future}.

In practice, for signals close to the  standard QCD band the possible displacement of a given signal will be obscured by the incertitude on the width of the band itself. 
 For experiments measuring directly the coupling to gluons, there is a large uncertainty on the theoretical prediction of $n$EDM$(f_a)$~\cite{JacksonKimball:2017elr}. For axion-photon searches, the experimental measurements are very precise with uncertainties stemming instead from the UV model dependence, i.e. the different matter content of axion models and different $E_k/\N_k$ factors. In summary, the smoking gun may turn out to be the multiplicity of signals, as experiments have reached impressive precision on frequencies.
 
 Furthermore, most of the experiments with best reach rely on the assumption that axions account for the dark matter (DM) of the universe. Haloscopes are sensitive to $\sqrt{\rho_{{\rm DM},local}}\times g_{aXX}$,  where $\rho_{{\rm DM},local}$ denotes the local DM density and $g_{aXX}$ a generic axion coupling to the visible world. Uncertainties stem from  $\rho_{{\rm DM},local}$  (which could be largely modified by $e.g.$ miniclusters, focusing effects, etc.) and, for the multiple QCD axion under discussion, on how DM is distributed among them.  If democratically distributed among the $N$ axion eigenstates, 
 the flux of each species would diminish by a factor of $N$ and consequently the bounds would be weakened.  We have illustrated this weakening of the projected bounds for CASPEr-electric  
  in Fig.~\ref{fig:fa_ma_example}, for the $N=3$ scenario.
   Nevertheless, the thermal evolution of the Universe with a multiple QCD axion may lead to very different spectra and evolution scenarios (e.g.~\cite{Kitajima:2014xla,Ho:2018qur,Cyncynates:2021xzw,Adams:2022pbo}), which deserve a dedicated future study.

The quest to unravel the fundamental symmetries of the visible and dark sectors of the Universe is inspiring. We have proven that the PQ solution to the strong CP problem within pure QCD leads in all generality to multiple axion signals  with well delimitated properties. This stems  from lifting the requirement   that the basis of axion-gluon interactions and the axion mass basis are simultaneously diagonal.
 In this perspective, the usual single QCD axion is just one particular case of the landscape of solutions, corresponding to zero mixing. The results open novel experimental territory 
 and 
 blur the distinction between the search for the QCD axion and that for ALPs
 in a huge region of the parameter space.
\\[0.6em]

 {\bf Acknowledgments.}---%
We are indebted to Daniel Alvarez-Gavela for useful discussions and input. We also thank Aneesh Manohar and Benjamin Grinstein for interesting discussions.
This project has received funding /support from the European Union's Horizon 2020 research and innovation programme under the Marie Sklodowska -Curie grant agreement No 860881-HIDDeN, and under 
the Marie Sklodowska-Curie Staff Exchange  grant agreement No 101086085-ASYMMETRY. The work of M. R.  is supported by the Marie Sklodowska -Curie grant agreement No 860881-HIDDeN. 
The work of P.Q. is supported in part by the U.S. Department of Energy Grant No. DE-SC0009919. 
B. G. acknowledge as well partial financial support from the Spanish Research Agency (Agencia Estatal de Investigaci\'on) through the grant IFT Centro de Excelencia Severo Ochoa No CEX2020-001007-S and through the grant PID2019-108892RB-I00 funded by MCIN/AEI/ 10.13039/501100011033.
 B.G. thanks very much the Particle Physics group of the University of California San Diego, where part of this work was carried out. Likewise, P.~Q.\ thanks the Galileo-Galilei Institute for theoretical Physics in Florence (GGI) for their warm hospitality, where part of this work was carried out. We are also thankful to Anna Lewis for helping to improve the writing of the manuscript.

\onecolumngrid
\appendix

\newpage

\section{UV completion}
\label{App:UV}

We consider a double version of the KSVZ model~\cite{Kim:1979if,Shifman:1979if}, with two vector-like fermions $\Psi_{1,2}\sim (3,1,0)$ and two singlet complex scalars ${S_{1,2}\sim (1,1,0)}$, i.e. 
\begin{equation}
\mathcal{L}_{\rm UV} =  |\partial_\mu S_1|^2 + |\partial_\mu S_2|^2 + \overline{\Psi}_1 \text{i} \slashed{D} \Psi_1 + \overline{\Psi}_2 \text{i} \slashed{D} \Psi_2 - \left[y_1 \overline{\Psi}_1 \Psi_1 S_1 +  y_2 \overline{\Psi}_2 \Psi_2 S_2 + \text{h.c.}\right] - V(S_{1,2})\,.
\end{equation}
In the absence of $V(S_{1,2})$, the Lagrangian is invariant under two independent global (and QCD anomalous) $U(1)$ axial symmetries. If the potential $V(S_{1,2})$ breaks both spontaneously, the $S_i$ fields can be decomposed 
in terms of two radial and two angular modes,
\begin{equation}
S_{1,2} = \frac{1}{\sqrt{2}}\left(\hat f_{1,2} + \rho_{1,2}\right) e^{\text{i} \hat a_{1,2}/\hat f_{1,2}}\,.
\end{equation}
Let us assume for simplicity $\hat f_1 = \hat f_2 =\hat f$.  The corresponding eigenstates are then ${a_1 = \frac{1}{\sqrt{2}} \left(\hat a_1 + \hat a_2 \right)}$ and $a_2 = \frac{1}{\sqrt{2}} \left(\hat a_1 - \hat a_2 \right)$. The first plays the role of the canonical QCD axion, with $m_1^2 f_1^2 = \chi_{\rm QCD}$, while the second is massless and decouples from the QCD sector.

Nevertheless, the potential may mix both fields,  breaking explicitly the two $U(1)$ global symmetries down to just one PQ symmetry.  A simple example is given by 
\begin{equation}
\label{eq:PQbreaking}
V(S_{1,2}) = \lambda S_1^3 S_2 + \text{h.c.}\,,
\end{equation}
which leads at low energies to an effective potential which, expanded to second order, reads:
\begin{align}
V_{\rm eff} & = \frac{1}{2}\chi_{\rm QCD}  \left(\frac{\hat a_1 + \hat a_2}{\hat f} - \bar{\theta} \right)^2 + \frac{\lambda}{4} \hat f^4  \left( \frac{3 \hat a_1 + \hat a_2}{\hat f} \right)^2 \,.
\end{align}
Defining $r \equiv \lambda \hat f^4 /(2 \chi_{\rm QCD})$ and considering the  preferred basis of Eq.~(\ref{Eq:MassMatricesPreferredbasis}), 
the corresponding mass matrix is 
\begin{equation} 
\label{2KSVZmatrix}
\mathbf{M}^2 = \frac{\chi_{\rm QCD}}{{\hat f}^2}
\begin{pmatrix}
2 + 8 r & -4 r \\
-4 r & 2 r
\end{pmatrix}\,,
\end{equation}
which indeed satisfies  the condition for PQ invariant potentials in \cref{Eq:PQrelation-chi}, with $1/ F^2 = 2 /\hat f^2$. The system thus reabsorbs $\bar \theta$ at the minimum, solving the strong CP problem.

Among the family of solutions in Eq.~(\ref{2KSVZmatrix}), it is possible to find maxion ones. Indeed, the maxion condition in Eq.~(\ref{eq:mTr}) is satisfied in this example  
for $r = 1/5$: two maxions are then obtained with $g_1 = g_2 = 2$. 

Note that many other explicit $V(S_{1,2})$ potentials can break explicitly the two $U(1)$ symmetries to one $PQ$ symmetry. For instance, $ V(S_{1,2}) \sim S_2^4$ would lead precisely to the effective potential in the toy model of~\cref{sec:N2}.

\section{A physical interpretation of the sum rule}
 \label{sec:new_physical_derivation}
 In \cref{sec:qcd_axion_likeness} we provided a rather mathematical proof of the sum rule. 
  In order to obtain a more physical intuition for the meaning of the \emph{axionness}, ${\beta_i\equiv}1/g_i$, and its constraint, we provide here an alternative derivation. The main idea is to impose that the modification of the Lagrangian under a PQ transformation coincides 
  in any starting basis and in the mass basis. This ``PQ matching'' constrains the possible values of the couplings to gluons and masses of the axion mass eigenstates leading to our sum rule, $\sum_i {\beta_i}=1$.

 On top of the mass eigenstates $\{a_{i}\}$, there are two linear combinations of particular importance for the QCD axion system. One of them is the combination that couples to the topological term, $\agg$, and the other is the one that transforms under the PQ symmetry, $a_{\text{PQ}}$. The latter is the combination whose shift $a_{\text{PQ}}\to a_{\text{PQ}} + \alpha  f_{\text{PQ}}$ is only broken by the QCD anomaly ({an equivalent definition is that it corresponds to the eigenvector of $\Mbf_B^2$ with vanishing eigenvalue}). We will now show how the \emph{axionness} of an axion will depend on the projections of a given mass eigenstate $a_i$ onto both of these linear combinations, $\agg$ and $\aPQ$. 

Let {us} consider a basis where the mass matrix stemming from the extra potential is diagonal, $\left\{\aPQ, \tilde{a}_l\right\}$ for ${l=1, \ldots N-1}$. Since the only term that breaks $U(1)_{{PQ}}$ is the $\aPQ \, G \tilde{G}$ term, the extra potential does not contain the $\aPQ$ field,
\begin{align*}
\mathcal{L}=\frac{\alpha_s}{8 \pi} {\underbrace{\left(\frac{\aPQ}{\fPQ}+ \sum_{l=1}^{N-1} \frac{\tilde{a}_l}{\hat{f}_l}\right)}_{\agg/F} G \tilde{G}-V_B\left(\tilde{a}_1, \ldots, \tilde{a}_{N-1}\right)}\,.
\label{Eq: PQ basis GGdual}
\end{align*}
(Note that $f_{\rm PQ}$, as defined above, does not need to coincide with the vacuum expectation value of the PQ scalar.)

The mass Lagrangian of the axion system 
reads,
\begin{align}
\mathcal{L}_{\hat \Mbf^2}={-}\frac{\chiq}{2}\left(\frac{\agg}{F}\right)^2 {-} V_B\left(\tilde{a}_{1,} \ldots \tilde{a}_{N-1}\right)={-} \frac{\chiq}{2}\left(\frac{\aPQ}{\fPQ}+\sum^{N-1} \frac{\tilde a_l}{\tilde f_l}\right)^2{-}V_B\left(\tilde{a}_{1,} \ldots \tilde{a}_{N-1}\right)\,,
\end{align}
which under a $\UPQ$ shift transforms as\footnote{Recall that, by definition, the relevant projections take the following values, $\left\langle \aPQ \mid \agg\right\rangle={F}/{\fPQ}$ and ${\left\langle a_i \mid \agg\right\rangle={F}/{f_i}}$.}
\begin{align*}
 \UPQ: \quad &\aPQ \longrightarrow \aPQ+\alpha \fPQ\,; \\
 &\agg \longrightarrow \agg+\alpha \fPQ \left\langle \agg \mid \aPQ\right\rangle= \agg+\alpha F \nn\,;  \\
& \mathcal{L}_{\hat \Mbf^2} \longrightarrow \mathcal{L}_{\hat \Mbf^2}^{\prime} \simeq - \frac{\chiq}{2}\left(\frac{\agg}{F}+\alpha\right)^2 \simeq -\frac{\chiq}{2} \frac{\agg^2}{F^2}-\alpha \chiq \frac{\agg}{F}-\frac{\alpha^2}{2} \chiq\,.
\end{align*}

Now let~{us} consider the diagonalized mass Lagrangian,
\begin{align}
\mathcal{L}_{ \Mbf^2}={-}\frac{1}{2} \sum_i^N m_i^2 a_i^2\,,
\end{align}
and perform the same $\UPQ$ transformation:
\begin{align}
 \UPQ: \quad &\aPQ \longrightarrow \aPQ+\alpha \fPQ \\
 &a_i \quad\longrightarrow a_i+\alpha \fPQ \left\langle a_i \mid \aPQ\right\rangle \nn\,;  \\
 &\mathcal{L}_{ \Mbf^2} \longrightarrow \mathcal{L}_{ \Mbf^2}^{\prime}  = -\frac{1}{2} \sum_i^N m_i^2\left(a_i+\left\langle a_i \mid \aPQ\right\rangle \alpha \fPQ\right)^2\,;  \\ 
 &\,\qquad\qquad\qquad=-\frac{1}{2} \sum_i^N m_i^2 a_i^2-\alpha \fPQ \sum_i^N m_i^2\left\langle a_i \mid \aPQ\right\rangle a_i  -\frac{1}{2} \alpha^2 \fPQ^2 \sum_i^N m_i^2 \left\langle a_i \mid \aPQ\right\rangle^2\nn \,.
\end{align}

{By imposing} the ``PQ matching" order by order,~we find:
\begin{align}
\O(\alpha):\qquad  \chi_{\mathrm{QCD}} \frac{{a}_{G \widetilde{G}}}{F}= \fPQ \sum_i^N m_i^2\left\langle a_i \mid \aPQ\right\rangle a_i\,,
\end{align}
and since $\agg=\sum\left\langle \agg \mid a_i\right\rangle a_i=F \sum a_i/f_i$, we conclude that
\begin{align}
\left\langle a_i \mid \aPQ \right\rangle=\frac{\chiq}{f_i \fPQ m_i^2}\,.
\end{align}

Finally, we reproduce our relation by noting that
\begin{align*}
1=\frac{\left\langle \aPQ \mid \agg\right\rangle}{\left\langle \aPQ \mid \agg\right\rangle}  =\sum_i^{{N}} \frac{\left\langle \aPQ \mid a_i\right\rangle\left\langle a_i \mid \agg\right\rangle}{\left\langle \aPQ \mid \agg\right\rangle}
 =\sum_i^N \frac{\chiq}{f_i \fPQ m_i^2} \frac{F}{f_i} \frac{\fPQ}{F}=\sum_i^N \frac{\chiq}{m_i^2 f_i^2}=\sum_i^N \frac{1}{g_i}\,,
\end{align*}
that is, 
\begin{align}
      \label{eq:betaPHY}
      \beta_i=\frac{1}{g_i}=\frac{\langle \aPQ \mid a_i\rangle\left\langle a_i \mid \agg\right\rangle}{\left\langle \aPQ \mid \agg\right\rangle}\,.
      \end{align} 
This shows  that for an axion mass eigenstate to coincide with the traditional single QCD axion, $\beta_i=1$, the overlap with both $\aPQ$ and $\agg$ needs to be maximal. Conversely, for a given physical axion to decouple from the sum rule, $\beta_i=0,$ it is necessary to either have a vanishing projection on the anomalous axion linear combination, $\langle a_i \mid \agg\rangle=0$, or on the combination implementing the PQ symmetry, $\left\langle \aPQ \mid a_i\right\rangle=0$. Indeed, it is these two vanishing projections that explain the decoupling of one axion from the sum rule in the $N=2$ example of~\cref{sec:N2}, in the limits $r\to 0$ and $r\to \infty$.

Besides proving the sum rule in Eq.~(\ref{eq:sum-rule}) without appealing to the eigenvalue-eigenvector theorem, this derivation directly suggests further relations such as 
      \begin{align}
      1= \sum_i^N\left\langle a_i \mid \aPQ \right\rangle=\frac{\chiq}{\fPQ}\sum_i^N\frac{1}{f_i \, m_i^2   }\,,
      \end{align}
      which may have a more limited phenomenological impact as compared to the previous sum rule, though, because we do not have direct experimental access to $\fPQ$. In an eventual multi-axion detection,  it could help to reconstruct the whole axion system. 
      
      \section{Decoupling from the sum rule and mass scales in the extra potential}\label{app:Vscale} 
 
As we explicitly derived for the toy model in ~\cref{sec:N2}, in the limits where the extra potential is either much smaller or much larger than the QCD contribution, one of the axions decouples from the sum rule. Two pertinent questions arise: how does this behaviour generalize for arbitrary $N$? Is the contribution from the extra potential required to be of the same order of magnitude as the QCD contribution, 
in order for all axions to play a relevant role in the sum rule?

 Let us consider the basis where the extra mass contribution is diagonal $\Mbf_B^2=\text{diag}(\tilde \lambda_1, \dots, \tilde \lambda_N)$, which corresponds to the states $\left\{\aPQ, \tilde{a}_1,\dots, \tilde{a}_{N-1} \right\}$.~This is the same basis used in the demonstration in App.~\ref{sec:new_physical_derivation}.

Using \cref{Eq: PQ basis GGdual}, the $g_i$-factors can be expressed as
\begin{align}
\label{gi}
g_i=\frac{m_{i}^2 \, F^2}{\left| \langle \agg|a_i\rangle\right|^2\chiq}
=\frac{m_{i}^2 \, }{\left| \langle \aPQ|a_i\rangle/\fPQ +\sum_j^{N-1} \langle\tilde a_j|a_i\rangle/\tilde f_j\right|^2\chiq}\,. 
\end{align}

For $\tilde \lambda_j \gg \chi_{\rm QCD}/F^2$, one of the mass eigenstates of the full matrix corresponds to $a_j\simeq \tilde a_j$, with mass $m_j^2\simeq \tilde \lambda_j$, and it is orthogonal to all the other $\tilde a_{k\neq j}$. Thus, its \emph{axionness} reads
  \begin{align}
 \frac{1}{g_j}\simeq \frac{\left| \langle \agg|\tilde a_j\rangle\right|^2\chiq}{\tilde \lambda_{j} \, F^2 \, }=  \frac{ (F/\tilde f_j)^2\, \chiq}{\tilde \lambda_{j} \, F^2}\leq  \frac{ \chiq}{\tilde \lambda_{j} \, F^2}\longrightarrow 0\,,
 \end{align}
  showing that this state effectively decouples from the sum rule.

  In the opposite limit, $\tilde \lambda_j \ll \chi_{\rm QCD}/F^2$, or equivalently $\varepsilon\equiv \tilde \lambda_j F^2/ \chi_{\rm QCD}\to 0$, one can show that the linear combination,
  \begin{align}
  a_\varepsilon = \frac{\aPQ}{\tilde f_j}- \frac{\tilde a_j}{\fPQ} + \O (\varepsilon)
  \end{align}
  corresponds to a mass eigenstate of the full matrix $\Mbf^2$ with eigenvalue $m_\epsilon^2\sim \tilde \lambda_j=  \varepsilon\,\chi_{\rm QCD}/F^2$. ~Therefore,
  \begin{align}
 \frac{1}{g_j}\simeq 
 \frac{\left| \langle \agg|\tilde a_\varepsilon\rangle\right|^2\chiq}{\tilde \lambda_{j} \,F^2 }
 \sim  \frac{\varepsilon^2}{\varepsilon}\longrightarrow 0\,,
 \end{align}
showing that this state also decouples from the sum rule.
  
Overall, we have demonstrated that whenever one eigenvalue of $M_B^2$, $\tilde \lambda_j$, is either much larger or much smaller than the QCD induced mass scale, $ \chi_{\rm QCD}/F^2$, one axion eigenstate decouples from the sum rule and the phenomenology is well described by the $N-1$ axion system. 

This statement depends however on the experimental precision.
For example, the matrix 
\begin{equation}
\label{eq:nmaxionN3}
\hat{\mathbf{M}}^2 = \frac{\chi_{\rm QCD}}{\hat{f}^2}\begin{pmatrix}
1 & 1 &1 \\
1 & 1 + \tilde{\lambda}_1 & 1 \\
1 & 1 & 1+ \tilde{\lambda}_2
\end{pmatrix}\,, 
\end{equation}
with $\tilde{\lambda}_1 = 10^{-3} \tilde{\lambda}_2 = 0.5$, generates a multiple axion system with $(g_1,\,g_2,\,g_3)\approx (1.2,\,7.3,\,497)$. Measuring $g_1$ and $g_2$ with sufficiently high precision could allow us to infer the existence of the third axion (or more) even though $1/g_3\ll 1$.

\section{Proof of the QCD maxion conditions}
\label{sec:qcd_maxion_condition}

We will show that the necessary and sufficient condition to generate QCD maxions is
\begin{align}
   \mathcal{B}_{n-k}^{\Mbf^2}=N\frac{\chiq}{F^2}\,  \mathcal{B}^{\Mbf_1^2}_{n-1-k}\,,
   \label{Eq: ConditionMaxion Bell}
\end{align}
determining the characteristic polynomials of $\Mbf^2$ and the minor $\Mbf_1^2$, 
    \begin{align}
    \mbox{\large $p$}_{{\,\Mbf^2}}(\lambda)=\sum_{k=0}^N \frac{(-1)^{N-k}}{(n-k) !}\, \mathcal{B}_{n-k}^{\Mbf^2}\,  \lambda^k  \qquad \text{and} \qquad
     \mbox{\large $p$}_{{\,\Mbf_1^2}}(\lambda)=\sum_{k=0}^{N-1} \frac{(-1)^{n-1-k}}{(n-1-k) !} \mathcal{B}^{\Mbf_1^2}_{n-1-k}\,  \lambda^k\,.
     \label{Wq:characteristic Polynomials}
    \end{align}
  
In order to prove this result, we will make use of the eigenvector-eigenvalue identity~\cite{JacobiDeBQ,Denton:2019pka}~{stated below}.\\[1pt]

    \underline{\sc Theorem}. If $A$ is an $N \times N$ hermitian matrix with eigenvalues $\lambda_1(A), \ldots, \lambda_{N}(A)$ and $i, j=1, \ldots, N$, then the $j^{\text {th }}$ component $v_{i j}$ of a unit eigenvector $v_i$ associated to $\lambda_i(A)$ is related to the eigenvalues $\lambda_1\left(M_j\right), \ldots, \lambda_{N-1}\left(M_j\right)$ of the minor $M_j$, obtained by removing the $j^{\text {th }}$ row and column~of A, by the formula
\begin{align}
\left|v_{i j}\right|^2 \prod_{k=1 ; k \neq i}^N\left(\lambda_i(A)-\lambda_k(A)\right)=\prod_{k=1}^{N-1}\left(\lambda_i(A)-\lambda_k\left(M_j\right)\right)\,.
\label{Eq:EigenvalueEigenvectorIdentity}
\end{align}

We will start by proving that $g_i=N,\,\forall i$ implies \cref{Eq: ConditionMaxion Bell}.
%
%
Let us first rewrite the $g$-factors in terms of the characteristic polynomials defined above,
\begin{align}
   g_i=
\frac{F^2}{\chiq }
\frac{\lambda_i \prod_{i\neq k}(\lambda_i-\lambda_k)}{\, \prod_{i\neq k}(\lambda_i-\lambda_k(M_1))}
=\frac{F^2}{\chiq }
\lim _{\lambda \rightarrow \lambda_i} \frac{\lambda \, \pM(\lambda)}{(\lambda-\lambda_i) \pMone(\lambda)} \,.
\end{align}
Since $\lambda_i$ is a zero of ${p_\mathbf{M}^2}\,(\lambda)$, it follows that ${p_\mathbf{M}^2}\left(\lambda_i\right)=0$ and thus~the $g$-factor can be expressed in terms of the derivative of the characteristic polynomial evaluated in $\lambda_i$,
\begin{align}
 \lim _{\lambda \rightarrow \lambda i} \frac{\pM(\lambda)}{\lambda-\lambda_i}= \lim _{\lambda \rightarrow \lambda_i} \frac{\pM(\lambda)-\pM\left(\lambda_i\right)}{\lambda-\lambda_i}=\left. \frac{d \pM(\lambda)}{d \lambda}\right|_{\lambda=\lambda_i}\quad \implies \quad 
 g_i
=\frac{F^2}{\chiq }  
\frac{\lambda_i }{ \pMone(\lambda_i)} 
\left.\frac{d \pM(\lambda)}{d \lambda}\right|_{\lambda=\lambda_i}
\label{Eq:gfactorDerivative}\,.
\end{align}

Differentiating the characteristic polynomial in \cref{Wq:characteristic Polynomials} and assuming the condition $g_i=N,\,\forall i $~we find:
\begin{align}
N \frac{\chiq }{F^2} \sum_{k=0}^{N-1} \frac{(-1)^{n-k-1}}{(n-k-1) !} \mathcal{B}_{n-k-1}^{\Mbf_1^2} \lambda_i^k=\sum_{k=0}^{N-1} \frac{(-1)^{n-k-1}}{(n-k-1) !} \mathcal{B}_{n-k}^{\Mbf^2} \lambda_i^k \\
{\Leftrightarrow}\sum_{k=0}^{N-1} \frac{(-1)^{n-k-1}}{(n-k-1) !} \left[N \frac{\chiq }{F^2} \mathcal{B}_{n-k-1}^{\Mbf_1^2}-\mathcal{B}_{n-k}^{\Mbf^2}\right] \lambda_i^k  {\,\equiv\,} A_{ik}C_k=0\,,
\label{Eq:expresionwithVandermonde}
\end{align}
where $A_{ik}\equiv \lambda_i^k$  is the Vandermonde matrix of dimension $N-1$, whose determinant is
\begin{align}
\operatorname{det}(A)=\prod_{0 \leq i<j \leq n}\left(\lambda_j-\lambda_i\right)\,.
\end{align}
Provided none of the eigenvalues are degenerate, $C_k=0$, proving the claim.

{The proof in the other direction, that is showing that the relations~(\ref{Eq: ConditionMaxion Bell}) imply $g_i=N\,\,\forall i $, can be trivially obtained by inserting these relations in \cref{Eq:gfactorDerivative}.}

\section{Comparison with clockwork scenario}\label{app:clockwork}

In general, clockwork matrices~\cite{Kim:2004rp,Choi:2014rja,Kaplan:2015fuy,Giudice:2016yja} do not generate maxions (one exception being the model comprising only 2 scalars). The reason being that the next neighbor interactions of clockwork scenarios are engineered to generate exponentially small mixings whereas {the maxion solutions} require sizable mixings. To see a concrete example, we focus on a scenario with three scalars, where the typical clockwork mass matrix (including the QCD contribution) reads:
\begin{equation}
\hat{\mathbf{M}}^2 =\frac{\chi_{\rm QCD}}{\hat{f}^2}
\begin{pmatrix}
0 &0 & 0 \\
0 & 0 & 0 \\
0 & 0 &1
\end{pmatrix} + r \frac{\chi_{\rm QCD}}{\hat{f}^2}
\begin{pmatrix}
1 & -q & 0 \\
-q & 1 + q^2 & -q \\
0 & -q & q^2 
\end{pmatrix}\,,
\end{equation}
with $q=3$ and assuming that only one field in the 3rd-site develops couplings to gluons.

One can easily check that the PQ condition ${\det \Mbf^2}/{\det \Mbf_1^2}={\chiq}/{{ F}^2}$ is satisfied {by this matrix}, as $F^2 = \hat{f}^2$.
To prove that this model does not generate maxions, it  suffices to show that the remaining two maxions conditions spanned by \cref{Eq: ConditionMaxion Bell} cannot be satisfied for the same $r$. Indeed,
\begin{align}
{\text{Tr} \,\Mbf^2}=N\, \frac{\chiq}{{F}^2} &\Leftrightarrow r =\frac{1}{10}\,, \\
{\text{Tr}^2 \,\Mbf^2} - {\text{Tr} \,\Mbf^2\cdot \Mbf^2}=N\, \frac{\chiq}{{F}^2} \text{Tr} \,\Mbf_1^2 &\Leftrightarrow r =0 \lor r=\frac{11}{182}\,.
\end{align}

\bibliographystyle{utphys}
\bibliography{draft}

\end{document}